\documentclass[useAMS,usenatbib]{mn2e}
\pdfoutput=1
\usepackage{aas_macros}
\usepackage{epsfig}
\usepackage{amsmath}
\usepackage{amssymb}
\usepackage{leftidx}
\usepackage{natbib}
\usepackage{xcolor}
\usepackage{hyperref}
\def\beq{\begin{equation}}
\def\eeq{\end{equation}}
\def\ba{\begin{eqnarray}}
\def\ea{\end{eqnarray}}
\def\bal{\begin{align}}
\def\eal{\end{align}}
\def\bxi{{\mbox{\boldmath $\xi$}}}

\begin{document}

\title[Collapsing Neutron Stars] {Dark Matter-induced Collapse of Neutron Stars:\\ A Possible Link Between Fast Radio Bursts and the Missing Pulsar Problem}

\author[Fuller \& Ott]{Jim Fuller$^{1,2}$,\thanks{Email: jfuller@caltech.edu}
Christian D. Ott$^1$,\\
\\$^1$TAPIR, Walter Burke Institute for Theoretical Physics, Mailcode 350-17, California Institute of Technology, Pasadena, CA 91125, USA
\\$^2$Kavli Institute for Theoretical Physics, Kohn Hall, University of California, Santa Barbara, CA 93106, USA
}

\label{firstpage}
\maketitle

\begin{abstract}

Fast radio bursts (FRBs) are an emerging class of short and bright radio transients whose sources remain enigmatic. Within the galactic center, the non-detection of pulsars within the inner $\sim \!10\,{\rm pc}$ has created a missing pulsar problem that has intensified with time. With all reserve, we advance the notion that the two
problems could be linked by a common solution: the collapse of neutron
stars (NS) due to capture and sedimentation of dark matter (DM) within
their cores. Bramante \& Linden (2014), Phys.\ Rev.\ Lett.~19, 191301
showed that certain DM properties allow for rapid NS collapse within
the high DM density environments near galactic centers while
permitting NS survival elsewhere. Each DM-induced collapse could
generate an FRB as the NS magnetosphere is suddenly expelled. This
scenario could explain several features of FRBs: their short time
scales, large energies, locally produced scattering tails, and high
event rates. We predict that FRBs are localized to galactic
centers, and that our own galactic center harbors a large population
of NS-mass ($M\sim1.4 M_\odot$) black holes. The DM-induced collapse
scenario is intrinsically unlikely because it can only occur in a
small region of allowable DM parameter space. However, if observed to
occur, it would place tight constraints on DM properties.

\end{abstract}


\section{Introduction}
\label{intro}

\subsection{Fast Radio Bursts}

Fast radio bursts (FRB) are a recently discovered class of radio transients \citep{lorimer:07,keane:12,thornton:13,burke-spolaor:14,spitler:14b} shrouded in both mystery and controversy. They are characterized by short ($t \sim {\rm ms}$) isolated radio bursts detected at $\sim{\rm GHz}$ frequencies with large dispersion measures ($D_M \sim {\rm several}\times 100\,{\rm cm}^{-3}{\rm pc}$).
The estimated event rate of detectable FRBs is large, up to $10^4$ per sky per day \citep{thornton:13}.

Unless FRBs are a form of local radio interference, they likely originate at cosmological distances (\citealt{luan:14,dennison:14,katz:14a,katz:14b,tuntsov:14}). If their observed $D_M$ originates from the intergalactic medium, typical FRB distances are large, $D \sim 2 \, {\rm Gpc}$) and emitted radio energies are high ($10^{38}$ to $4 \times 10^{40}\,{\rm erg}$) (\citealt{thornton:13,kulkarni:14,dolag:14}). Conversely, the scattering tails observed in some FRBs cannot be explained by scattering by the IGM, and likely originates within the host galaxy of the source \citep{luan:14,katz:14a}. This may indicate that at least some FRBs originate near galactic centers. Typical comoving distances of $D \! \sim \! 2\,{\rm Gpc}$ and an event rate of $10^4\,{\rm day}^{-1}$ imply a volumetric rate of $R \sim 10^{-4}\,{\rm yr}^{-1}{\rm Mpc}^{-3}$, comparable to the volumetric core-collapse supernova rate \citep{li:11b}. Many progenitors have been suggested to be the sources of FRBs, although most seem unlikely (see the discussion in \citealt{kulkarni:14}). Here we simply re-iterate that the short durations, large luminosities, and high brightness temperatures naturally suggest a neutron star (NS) origin. 


\subsection{Missing Galactic Center Pulsar Problem}

Within our own galaxy, the missing pulsar problem at the galactic center (GC) has recently become more puzzling. Despite several deep searches \citep{johnston:06,deneva:09,macquart:10,bates:11}, no ordinary pulsars have been discovered within a projected distance of $10\,{\rm pc}$ of the GC despite predictions that there should be more than $10^3$ active radio pulsars in this region \citep{pfahl:04}. The problem recently intensified with the radio detection of a magnetar within the central parsec \citep{eatough:13}. Subsequent studies \citep{bower:14,spitler:14a} indicated that the magnetar indeed lies very close to Sgr A* and that the temporal scattering was not severe enough to have prevented the detection of ordinary pulsars in previous surveys. 

It remains possible that (for some unknown reason) ordinary pulsars near the GC
have a steep spectral index and are not bright enough at high
frequencies to be detected with current instrumentation
\citep{bates:11}. Alternatively, \cite{dexter:14} have suggested that
the missing pulsar problem is created by a very high magnetar
formation efficiency near the GC, such that GC pulsars spin down so
rapidly that they are unlikely to be observed before crossing the
death line. 
Given our current (incomplete) understanding of magnetar formation (e.g.,
\citealt{thompson:93,keane:08}), there is no known physical process or
condition that could favor magnetar formation near the GC.



\subsection{Dark Matter-Induced Neutron Star Collapse}

We propose that the seemingly separate problems posed by FRBs and
missing pulsars may have the same solution: the DM-induced collapse of
neutron stars (NS). In this scenario (recently described by
\citealt{bramante:14}, see also
\citealt{kouvaris:11b,kouvaris:12,mcdermott:12,bramante:13} and references
therein), ambient DM particles scatter off of nucleons within an NS
and become gravitationally bound. The DM particles continue to scatter
until they thermalize at the NS temperature and sink to the center of the NS where they accumulate within a
small volume. When the DM agglomeration exceeds a critical mass, the DM forms a black hole (BH)
at the NS center that may consume the NS. The NS collapses into the BH on a dynamical time scale ($t_{\rm dyn}
\sim 1\,{\rm ms}$), and a $\sim \!1.4 M_\odot$ BH remnant is left
behind.


During the collapse, the magnetosphere must suddenly detach from the BH horizon, and
the rapid reconfiguration of the field lines generates a burst of electromagnetic
energy. If a small fraction of this energy is radiated
in radio wavelengths, an FRB is generated. The process occurs most
rapidly near the GC where DM densities are highest, causing NSs near
the GC to quickly collapse (but permitting long-term NS survival in
the solar neighborhood and in globular clusters), thereby accounting
for the lack of pulsar detections near the GC.

There is only a limited range in DM properties that allow for the process described above to occur. These have been explored in the works listed above, here we simply restate their basic results. First, there is only a narrow (2-3 orders of magnitude) range of DM scattering cross sections $\sigma$ that allows for rapid DM accumulation within GC NSs while allowing more local NSs to survive for several Gyr. Second, the DM must be weakly annihilating, otherwise it will annihilate upon accumulation at the center of the NS and will not be able to form a BH. Third, the DM mass, particle type (boson vs. fermion), self interaction, and NS temperature must allow for DM sedimentation and BH formation at the center of the NS. 


The DM-collapse scenario also hinges on the DM density profiles of NS
host galaxies. The DM capture rate is proportional to the ambient DM
density, $ \rho_{DM}$, and so the NS life time scales as $t_c \propto
\rho_{\rm DM}^{-1}$. 
To collapse a GC pulsar located at a galacto-centric radius $r=1\,{\rm
  pc}$ within $10^{6} \, {\rm yr}$, while allowing for the $10^{10} \,
{\rm yr}$ survival of a solar neighborhood NS ($r \sim 10^4 {\rm
  pc}$), requires the DM density within the inner parsec to be
enhanced by at least a factor of $10^4$.


\section{Generation of a Fast Radio Burst}
\label{frbgen}

The detailed electromagnetic signature of a collapsing NS is well beyond the scope of this paper, but here we make some rough estimates. The FRB emission energy cannot greatly exceed the energy contained in the magnetosphere,
\beq
\label{etot}
E_M \sim 10^{42} \, {\rm erg} \, B_{12}^2,
\eeq
where $B_{12}$ is the dipole field strength in units of $10^{12}\,{\rm G}$.
The radio emission from our collapsing NS scenario is similar to that
described in \cite{falcke:14} (FR14), who examined rapidly-rotating supramassive NS collapse.

During the collapse, electrons/positrons bound to field lines generate coherent radiation as the field reconfigures. If $\sim \! \! 10^{-3}$ of the field energy is emitted as radiation near GHz frequencies, an FRB can be generated. However, the coherent curvature radiation will initially have frequency near $\nu \! \sim \! 1/t_{\rm dyn} \! \sim \! {\rm kHz}$. To emerge at GHz frequencies, the radiation must be up-converted to higher frequencies. FR14 argue that this will occur until the coherent radiation is near the plasma frequency
\beq
\label{nuplasma}
\nu_p \sim 2\,{\rm GHz} \  B_{12}^{1/2} P_1^{-1/2} \,, 
\eeq
where $B_{12}$ is the field strength in units of $10^{12} \, {\rm G}$, and $P_1$ is the spin period in seconds. Following their analysis, we calculate that electrons are boosted to Lorentz factors of $\gamma \sim 60 \, B_{12}^{1/6} P_1^{-1/2}$, and radiate a power near GHz frequencies of
\beq
\label{powerrad}
P_R \sim 2 \times 10^{41}\,{\rm erg}\,{\rm s}^{-1}\, B_{12}^{13/6} P_1^{-13/6}.
\eeq
We expect the emission to be generated over roughly one magnetosphere light crossing time, $t \sim 1 \, {\rm ms}$. The corresponding radio energy emission is then $E_R \sim 2 \times 10^{38}\,{\rm erg}$, but can be substantially larger for higher field strengths or smaller rotation periods. While the frequency of equation \ref{nuplasma} and the power estimate of equation \ref{powerrad} are encouraging, the details of the emission process may be quite complex, and these estimates may not be robust.

The radiative power of equation \ref{powerrad} can be compared to simulations of the non-rotating collapse of a NS by \cite{lehner:12} (see also \citealt{palenzuela:13,dionysopoulou:13}), who find $P_{\rm rad} \sim 10^{42} \! - \!10^{43} \, B_{12}^2 \, {\rm erg} \,{\rm s}^{-1}$ over a duration of roughly a millisecond, implying $E_{\rm rad} \sim 10^{39} \! - \! 10^{40} \, B_{12}^2 \,{\rm erg}$. 
Unfortunately, these authors do not compute a radiation spectrum, so it is unclear how much energy is emitted in the radio band. If a significant fraction of it is emitted near GHz frequencies, as argued by FR14, then an FRB could be generated. It is possible that some of the radiation is emitted at optical/UV/X-ray/$\gamma$-ray wavelengths, however, any transient produced is unlikely to be detected. Its maximum luminosity is $L \lesssim 10^{45} \, {\rm erg \, s}^{-1}$, and its duration is expected to last $t \lesssim 1 \, {\rm ms}$, preventing even very rapid follow-up observations. 

The energy estimates above are similar to the isotropic radio energies estimated by \cite{thornton:13} and \cite{spitler:14b}, but smaller than those estimated by \cite{kulkarni:14} and \cite{dolag:14}. The total energy emitted depends on the slope of the spectral emission, and we do not attempt to quantify those details here. We simply note that our rudimentary estimates yield a radio luminosity similar to that inferred for FRBs. Slightly beamed emission (as seen in \citealt{lehner:12}) would entail smaller total radiative luminosities, but larger collapse rates.


The observed scattering tails in some FRBs are also consistent with the DM-induced collapse scenario, as \cite{luan:14} and \cite{katz:14b} have shown that the scattering is most likely generated near the source deep within a host galaxy. Similarly, our scenario entails that FRBs occur near GCs where the scattering could be produced, e.g., by the ionized surfaces of nearby molecular clouds \citep{lazio:98}. Moreover, at least some of the observed $D_M$ may be contributed by the host galaxy. The NE2001 model for the electron density distribution \citep{cordes:02} implies an average number density of $n_e \sim 0.2\, {\rm cm}^{-3}$ within the inner 500 pc \citep{ferriere:07}. A typical FRB occurring within the inner 500 pc (see Section \ref{rates}) would then incur a dispersion measure of $D_M \sim 100 \, {\rm cm}^{-3} \, {\rm pc}$, depending on the viewing angle. An additional $\sim 50 \, {\rm cm}^{-3} \, {\rm pc}$ will likely be contributed by its host's galactic halo \citep{dolag:14}, entailing a net contribution of $D_M \approx 150 \, {\rm cm}^{-3} \, {\rm pc}$ from the host galaxy. If the FRB occurs within the inner $\sim 100 \, {\rm pc}$, the local $D_M$ could be much larger, on the order of  $D_M \sim 10^3 \, {\rm cm}^{-3} \, {\rm pc}$ \citep{cordes:02}. This implies FRBs could be somewhat closer (by $\sim 25 \%$ or more) than estimated by assuming the observed $D_M$ is generated in the IGM.

\section{Rates and Galactic Center Neutron Star Depletion}
\label{rates}

The NS collapse time is determined by the DM accumulation rate and the amount of DM required to form a BH at the NS center. We calculate both using the method outlined in \cite{bramante:14}. In general, only a very small amount of DM ($M_{\rm acc} \lesssim 10^{-10} M_\odot$) is accreted before collapse. The most salient detail of the process is that the collapse time scales as $t_c \propto \rho_{DM}^{-1}$, allowing for short collapse times in high DM density environments. 

\begin{figure}
\begin{center}
\includegraphics[scale=0.45]{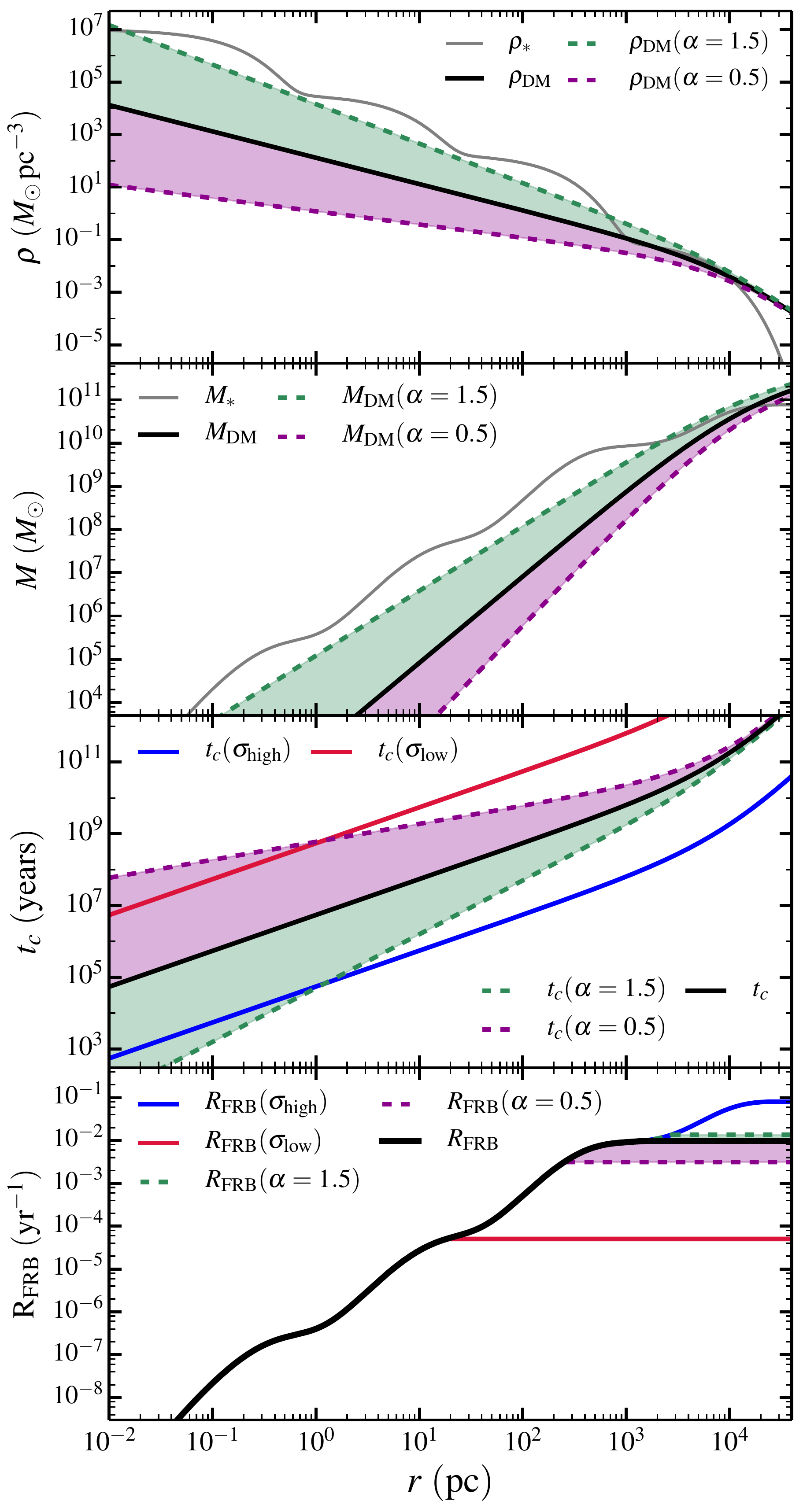}
\end{center} 
\caption{ \label{FRB} {\bf Top:} Spherically averaged density profile of a simple galactic model, including the stellar mass density $\rho_*$ and the DM density $\rho_{\rm DM}$. Green shaded regions have steep DM density profiles ($1 < \alpha < 1.5$), while purple shaded regions have shallow DM density profiles ($0.5 < \alpha < 1$). {\bf Second Row:} Corresponding enclosed mass for our galactic models. {\bf Third Row:} NS collapse time (black line) as a function of radius, for bosonic DM with a nucleon scattering cross section of $\sigma = 3 \times 10^{-49}\, {\rm cm}^{2}$ and mass $m = 10 \, {\rm MeV}/c^2$. Shaded regions correspond to the DM density profiles above. We have also plotted $t_c$ for $\alpha=1$ but with a DM cross section that is larger (blue) and smaller (red) by a factor of $10^2$. {\bf Bottom:} Enclosed FRB rates, assuming each NS collapse generates an FRB. }
\end{figure}

To estimate the FRB rate in a Milky Way-like galaxy, we construct a simple galactic model similar to the exponential spheroid model of \cite{sofue:13}. The model contains components from the central star cluster at $r<1\,{\rm pc}$, inner bulge at $r<20\,{\rm pc}$, bulge at $r<1\,{\rm kpc}$, and exponential disk at $r<20\,{\rm kpc}$. We also include a DM component with a generalized NFW profile \citep{navarro:97}, 
\beq
\rho(r) = \frac{\rho_0}{(r/R_s)^\alpha (1 + r/R_s)^{(3-\alpha)}}
\eeq
with scale radius $R_s=12\,{\rm kpc}$ and density normalization $\rho_0 = 1.1 \times 10^{-2} M_\odot \, {\rm pc}^{-3}$. These models have $\rho \propto r^{-\alpha}$ near GCs, with $\alpha=1$ for a standard NFW profile. For simplicity, we assume a constant DM velocity dispersion of $v_{\rm DM} = 200 \, {\rm km} \, {\rm s}^{-1}$ at all radii. Figure \ref{FRB} shows the density profile and enclosed mass of our model.

We caution that the DM profiles near the centers of galaxies are somewhat poorly constrained, especially within the inner $\sim \!\! 100 \, {\rm pc}$, which is typically not probed by observations nor simulations. The current understanding is that in galaxies less massive than the Milky Way, the DM profiles are cored due to star formation feedback, i.e., their central density profiles are shallower than the cuspy $\rho_{\rm DM} \propto r^{-1}$ profiles of an NFW profile \citep{delpopolo:09,inoue:11,maccio:12}. However, simulations of more massive galaxies indicate DM profiles with $\rho_{\rm DM} \propto r^{-1}$ \citep{dicintio:14}, or in many cases even steeper density profiles due to adiabatic contraction \citep{blumenthal:86,gnedin:04,gustafsson:06}. Lensing studies have also revealed steep central DM density profiles in massive early type galaxies \citep{tortora:10,sonnenfeld:12,grillo:12}, and high DM densities within central cluster galaxies \citep{newman:13}. Hence, we expect long (greater than a Hubble time) NS collapse times in low-mass galaxies, although short collapse times are achievable in massive spirals, early type galaxies, and central cluster galaxies, where central DM densities may greatly exceed $1 \, M_\odot \, {\rm pc}^{-3}$.  Below, we shall consider a range of DM profiles, with slopes $0.5 < \alpha < 1.5$.


As a matter of demonstration, we choose a non-self-interacting and non-annihilating bosonic DM particle with mass of $m = 10 \,{\rm MeV}/c^2$ and nucleon scattering cross section $\sigma = 3 \times 10^{-49}\,{\rm cm}^2$. For these choices, BH formation will occur upon formation of a super-Chandrasekhar mass Bose-Einstein condensate at the center of the NS, and the BS will be massive enough to consume the NS on a short timescale. For more massive DM particles, the number of DM particles required for BH formation will depend on the NS core temperature.
\footnote{An NS core temperature $T_c \! = \! 10^{4}\,{\rm K}$ is reasonable for old NSs in the photon-cooling phase (see, e.g., \citealt{yakovlev:04}, and cooling tracks obtained with the NSCool code; \url{http://www.astroscu.unam.mx/neutrones/NSCool}). Young NSs (with ages less than $\! \sim \!2 \times 10^6 \,{\rm yr}$) may be substantially warmer, which may raise the DM mass required for BH formation and therefore increase collapse times. This could prevent prompt NS collapse near GCs for massive ($m \gtrsim {\rm GeV}/c^2$) DM particles. Additionally, DM particles with $m \sim 1 \,{\rm GeV}/c^2$ may not be able to thermalize and settle within the NS fast enough to permit prompt collapse. We refer the reader to \cite{bramante:13} and \cite{bramante:14} for discussion of how NS temperature affects the DM accumulation process.}
Additional issues (e.g., self-interacting bosonic DM, fermionic DM, the DM thermalization time-scale, and BH evaporation) can affect the DM accumulation, sedimentation, and BH formation process. A full exploration of these issues is beyond the scope of this work, and we refer the reader to previous studies (e.g., \citealt{kouvaris:11b,kouvaris:12,mcdermott:12,bramante:13,bramante:14}. For our chosen parameters, the collapse time is dependent primarily on the DM cross section $\sigma$ and ambient DM density $\rho_{\rm DM}$, with the collapse time scaling as $t_c \propto (\rho_{\rm DM} \sigma)^{-1}$.

Using the DM properties and density profile described above, we compute the NS collapse time $t_c$ as a function of radius, shown in Figure \ref{FRB}. In the solar neighborhood, the collapse time is $t_c \sim 10^{11}\,{\rm yr}$, while in the central parsec, $t_c \lesssim \mathrm{few}\,\times 10^{6}\,{\rm yr}$. DM-induced collapse could thus substantially reduce the number of detectable radio pulsars within the central $\sim \! 10 \, {\rm pc}$ assuming typical radio pulsar lifetimes are $10^7 \! - \! 10^8 \, {\rm yr}$ \citep{faucher:06}. Steeper DM density profiles (green shaded area in Figure \ref{FRB}) will allow for shorter collapse times and more efficient pulsar destruction. Very steep profiles are problematic for the DM-induced collapse scenario because they predict very short collapse times that would preclude the observation of the GC magnetar, which is estimated to lie near $r \sim 0.1 \, {\rm pc}$ with an age of $\sim \! 10^4\,{\rm yr}$ \citep{eatough:13,bower:14,spitler:14a,dexter:14}. Shallower DM profiles (purple shaded area) would not allow for prompt NS collapse and could not resolve the missing pulsar problem, although they may still allow for relatively large FRB rates. A much smaller DM cross section (red line in Figure \ref{FRB}) would not allow for timely NS collapse and could not resolve the missing pulsar problem nor generate enough FRBs. A much larger DM cross section (blue line) would cause NSs to quickly collapse even outside of GCs, and can be ruled out by existing observations of old NSs.

To compute an approximate FRB rate, we assume a constant star formation rate and galaxy age of $t_G = 10^{10}\,{\rm yr}$, such that the star formation rate within a radius $r$ is simply $\dot{M}_* = M(r)/t_G$. Star formation rates peaking at early times will yield older populations of NSs and will increase the FRB rate.
Given a current star formation rate of $\dot{M}_{*,0} \approx 1.6 \, M_\odot \,{\rm yr}^{-1}$ \citep{licquia:14} and a galactic core-collapse supernova rate of $R_{\rm SN} \sim 2 \times 10^{-2}\,{\rm yr}^{-1}$ \citep{li:11b}, we calculate a NS creation rate of $\dot{M}_{\rm NS} \sim 1.5 \times 10^{-2} \dot{M}_*$, assuming each supernova generates a NS of mass $M_{\rm NS} \simeq 1.4 \, M_\odot$. Finally, we set the FRB rate per unit volume equal to
\begin{align}
\label{RFRB}
&\frac{ d R_{\rm FRB}}{dV} = \frac{1}{ M_{\rm NS} } \frac{d \dot{M}_{\rm NS}}{dV} & \quad {\rm if} \ t_c < t_G \,, \nonumber \\
&\frac{ d R_{\rm FRB}}{dV} = 0 & \quad {\rm if} \ t_c > t_G \,.
\end{align}
The FRB rate within a radius $r$, $R_{\rm FRB}$, is simply the volume integral over equation \ref{RFRB}, and is shown in the bottom panel Figure \ref{FRB}. This demonstrates that a Milky Way-like galaxy is capable of producing a total FRB rate of  $R_{\rm FRB} \sim 10^{-2}\,{\rm yr}^{-1}\,{\rm gal}^{-1}$. 

In comparison, the core-collapse supernova rate is $R_{\rm SN} \sim 2 \times 10^{-2}\,{\rm yr}^{-1}\,{\rm gal}^{-1}$ for a Milky-Way like galaxy \citep{li:11b}. In our simple model, only about one tenth of galactic NSs (those born within roughly 1 kpc of the GC) can collapse within a Hubble time. However, the current FRB rate is comparable to the current supernova rate because the average galactic past star formation rate ($\dot{M} \sim 10 \, M_\odot \, {\rm yr}^{-1}$) is larger than the current star formation rate. Given a volumetric core-collapse supernova rate (at $z=0$) of $R_{\rm SN} \sim 10^{-4} \, {\rm Mpc}^{-3} \, {\rm yr}^{-1}$ \citep{li:11b}, a plausible FRB rate is then
\beq
\label{rate}
R_{\rm FRB} \sim 5 \times 10^{-5} \, {\rm Mpc}^{-3} \, {\rm yr}^{-1} \,.
\eeq

The observed FRB rate is somewhat debated, but may be as large as $10^{4} \, {\rm day}^{-1}$ \citep{thornton:13}. This corresponds to a volumetric rate of
\beq
\label{Robs}
R_{\rm obs} \sim 10^{-4} \, {\rm Mpc}^{-3} \, {\rm yr}^{-1} \bigg( \frac{D}{2\,{\rm Gpc}} \bigg)^{-3} \, ,
\eeq
where $D$ is the comoving distance out to which most FRBs are observed.
The lack of FRB detections in lower galactic latitude surveys \citep{petroff:14a,burke-spolaor:14} indicates the FRB rate may be substantially lower than that quoted by \cite{thornton:13}. It is possible that the DM-induced collapse rate is dominated by other types of galaxies (e.g., early type galaxies, galaxies in clusters) with larger DM densities. We therefore argue that the DM-induced collapse rate may be high enough to account for most FRBs. 


\section{Discussion and Predictions}


There are several reasons why the DM-induced collapse scenario may not occur. The most likely is that DM properties are simply not compatible with this scenario, for the multitude of possibilities discussed in \cite{bramante:13}.
On the contrary, if DM-induced collapse is observed to occur, it would provide constraints on DM properties.
Below, we list several predictions of the DM-induced collapse scenario and its relation to the GC missing pulsar problem and FRB generation. If these predictions can be falsified, we can discount the scenario. However, if the predictions hold up, the scenario must be regarded seriously.

1. Dark matter masses, scattering cross sections, and self-interaction properties are in accordance with the limits discussed in \cite{bramante:14}.\footnote{Note that the allowable parameter space depends on DM self-interactions, and in certain cases on NS interior temperature. Therefore, the allowable parameter space shown in \cite{bramante:14} should be viewed as a guideline rather than a strict limit.}

2. There is no GC X-ray or $\gamma$-ray excess due to DM annihilation.

3. Large galaxies contain cusped DM density profiles, with central DM densities exceeding $\sim \! \! 1 \,{M_\odot}\,{\rm pc}^{-3}$. 

4. Long-lived NSs do not exist near most GCs, i.e., there should be no observed NSs with ages much larger than the collapse time shown in Figure \ref{FRB}. 
The discovery of an old millisecond pulsar or low mass NS X-ray binary within the inner $\sim \! 50 \,{\rm pc}$ of the GC would disfavor the NS-induced collapse scenario.

5. There exists a large population of NS-mass ($M \approx 1.4 \, M_\odot$) BHs near GCs. The discovery of a low mass X-ray binary containing a $M \lesssim 2 \, M_\odot$ BH near the GC would reinforce the DM-induced NS collapse scenario.

6. FRBs are localized near (within $\sim \! 1 \,{\rm kpc}$ of) GCs of massive galaxies or central cluster galaxies where DM densities are large. A magnetar flare FRB scenario (\citealt{kulkarni:14,lyubarsky:14,pen:15}) may entail that many FRBs occur near GCs, but also allows FRBs to arise from young stellar populations in dwarf galaxies. In contrast, dwarf galaxy hosts (where DM densities are low) are unlikely in the DM-induced collapse scenario.

7. FRBs exhibit little or no associated optical or X-ray transient, in agreement with the recent non-detection of any transient associated with FRB 140514 \citep{petroff:14b}. FRBs will not ever repeat, although multiple FRBs could arise from the same galaxy as long as its GC is DM-rich.

We hope the exciting implications of these scenarios will reinvigorate searches for FRBs and GC pulsars, and we anticipate that future observations will elucidate the nature of these enigmas.


\section*{Acknowledgments}

We thank Peter Goldreich, Phil Hopkins, Christine Moran, Evan
Kirby, and Sterl Phinney for useful discussions. JF acknowledges support from NSF under
grant no. AST-1205732 and through a Lee DuBridge Fellowship at
Caltech. CDO acknowledges support from NSG through grants PHY-1151197
and AST-1205732.


\bibliography{bibliography/jet_references,bibliography/bh_formation_references,bibliography/gw_references,bibliography/sn_theory_references,bibliography/grb_references,bibliography/nu_obs_references,bibliography/methods_references,bibliography/eos_references,bibliography/NSNS_NSBH_references,bibliography/stellarevolution_references,bibliography/nucleosynthesis_references,bibliography/gr_references,bibliography/sn_observation_references,bibliography/numrel_references,bibliography/gw_data_analysis_references,bibliography/ns_references,bibliography/stellar_oscillations_references,bibliography/gw_detector_references,bibliography/galactic_center_and_frb_references,bibliography/dark_matter_references,bibliography/sn_rates_references,bibliography/populations_references,bibliography/sgr_magnetar_references}

\end{document}